\documentclass[12pt]{article}
\usepackage[reqno]{amsmath}
\usepackage{bbm}
\usepackage{epsfig}
\usepackage{array}
\usepackage{float}

\usepackage{a4}

\usepackage{a4wide}
\usepackage{wasysym}

\def\be{\begin{equation}}
\def\ee{\end{equation}}
\def\gs{\mathrel{
   \rlap{\raise 0.511ex \hbox{$>$}}{\lower 0.511ex \hbox{$\sim$}}}}
\def\ls{\mathrel{
   \rlap{\raise 0.511ex \hbox{$<$}}{\lower 0.511ex \hbox{$\sim$}}}}

\newcommand{\ba}{\begin{array}{c}}
\newcommand{\baz}{\begin{array}{cc}}
\newcommand{\bad}{\begin{array}{ccc}}
\newcommand{\bea}{\begin{equation} \begin{array}{c}}
\newcommand{\eea}{ \end{array} \end{equation}}
\newcommand{\ea}{\end{array}}
\newcommand{\D}{\displaystyle}
\newcommand{\dms}{\mbox{$\Delta m^2_{\odot}$}}
\newcommand{\dma}{\mbox{$\Delta m^2_{\rm A}$}}
\newcommand{\meff}{\mbox{$\langle m \rangle$}}
\newcommand{\eV}{\mbox{ eV}}

\newcommand{\im}{i}

\newcommand{\diag}{\ensuremath{\mathrm{diag}}}

\newcommand{\ts}{\mbox{$ \tan^2 \theta_{\rm \odot}$}}

\begin{document}

\title{\vspace{-2cm}
\hfill {\small TUM--HEP--597/05}\\
\vspace{-0.3cm}
\hfill {\small hep--ph/0507300} 
\vskip 0.2cm
\bf Flavor Symmetry $L_\mu - L_\tau$ and quasi--degenerate Neutrinos
}
\author{
Werner Rodejohann\thanks{email: \tt werner$\_$rodejohann@ph.tum.de}~~
and~~
Michael A.\ Schmidt\thanks{email: \tt michael$\_$schmidt@ph.tum.de}~~
\\\\
{\normalsize \it Physik--Department, Technische Universit\"at M\"unchen,}\\
{\normalsize \it  James--Franck--Strasse, D--85748 Garching, Germany}
}
\date{}
\maketitle
\thispagestyle{empty}
\vspace{-0.8cm}
\begin{abstract}
 
\noindent 
Current data implies three simple forms of the neutrino mass matrix, 
each corresponding to the conservation of a non--standard lepton charge.   
While models based on $L_e$ and $L_e - L_\mu - L_\tau$ are well--known, 
little attention has been paid to $L_\mu - L_\tau$. 
A low energy mass matrix conserving $L_\mu - L_\tau$ implies 
quasi--degenerate light neutrinos. Moreover, it is 
$\mu$--$\tau$ symmetric and therefore 
(in contrast to $L_e$ and $L_e - L_\mu - L_\tau$) automatically 
predicts maximal atmospheric neutrino mixing and zero $U_{e3}$. 
A see--saw model based on $L_\mu - L_\tau$ is investigated and testable 
predictions for the neutrino mixing observables are given. 
Renormalization group running below 
and in between the see--saw scales is taken into account in our analysis, 
both numerically and analytically. 

\end{abstract}

\newpage

\section{Introduction} 
 
Impressive steps towards identifying the structure of the 
neutrino mass matrix $m_\nu$ have been taken 
in recent years \cite{revs}. In the charged lepton basis 
$m_\nu$ is diagonalized by the PMNS matrix \cite{PMNS} $U$ defined via 
\be \label{eq:def}
U^T \, m_\nu \, U = \diag(m_1,m_2,m_3)~. 
\ee
It can be parametrized as  
\begin{equation}
 U=\left(
 \begin{array}{ccc}
 c_{12} \, c_{13} & s_{12}\, c_{13} & s_{13}\, e^{-i \delta}\\
 -c_{23}\, s_{12}-s_{23}\, s_{13}\, c_{12}\, e^{i \delta} &
 c_{23}\, c_{12}-s_{23}\, s_{13}\, s_{12}\, e^{i \delta} & s_{23}\, c_{13}\\
 s_{23}\, s_{12}-\, c_{23}\, s_{13}\, c_{12}\, e^{i \delta} &
 -s_{23}\, c_{12}-c_{23}\, s_{13}\, s_{12}\, e^{i \delta} & c_{23}\, c_{13}
 \end{array}
 \right) \, P ~,
\end{equation}
where $P = \diag(e^{-i \varphi_1/2},e^{-i \varphi_2/2},1)$ contains  
the Majorana phases and  $c_{ij}$, $s_{ij}$ are 
defined as $\cos\theta_{ij}$ and $\sin\theta_{ij}$, respectively. 

\noindent 
The smallness of neutrino masses is commonly attributed to the 
see--saw mechanism \cite{see--saw}, which states that integrating out 
three heavy Majorana neutrinos with masses $M_{1,2,3}$ leads to a low energy 
neutrino mass matrix given by 
\be \label{eq:see--saw} 
m_\nu = - m_D \, M_R^{-1} \, m_D^T~. 
\ee
Here $M_R$ denotes the mass matrix of the three Majorana neutrinos 
$N_{1,2,3}$ and $m_D$ is a Dirac mass matrix resulting from 
the coupling of the Higgs doublet to the left--handed neutrinos and 
the $N_i$. 

\noindent 
On the other hand, the unexpected form of the mixing 
scheme implied by the data, 
namely one small ($\theta_{13}$) 
and two large ($\theta_{12,23}$ with $\theta_{12} \ls \theta_{23}$) 
angles in combination with  
a rather moderate mass hierarchy, poses a more difficult problem. 
It might hint towards a particular symmetry, which 
seems not to be present in the quark sector. 
In order to find this  
underlying symmetry of the neutrino mass matrix, one might consider 
minimal models and/or symmetries which are able to accommodate the data. 
For instance, one could introduce a non--standard flavor charge corresponding 
to an Abelian $U(1)$ symmetry and demand that the neutrino mass 
matrix conserves this flavor charge. This will  
imprint a particular form onto $m_\nu$ and lead to certain mixing scenarios 
which can then be compared to the data. 
For three active flavor states there are the charges $L_e$, $L_\mu$ and 
$L_\tau$, out of which one can construct 10 linearly  
independent combinations of the 
form $c_e \, L_e + c_\mu \, L_\mu + c_\tau \, L_\tau$ with 
$c_{e, \mu, \tau} = 0$ or $\pm1$.  
Neutrino data implies that these symmetries must be approximate. 
Nevertheless, already at the present stage we are 
capable of ruling out six of them, 
leaving us with four testable cases. Those four cases are 
easily to distinguish, because 
their predictions for the neutrino mass spectrum and for 
the effective mass governing neutrinoless double beta decay \cite{meff} 
are very different: 
\begin{itemize} 
\item conservation of $L_e$ implies the following form for the neutrino 
mass matrix \cite{le}: 
\be \label{eq:le}
m_\nu = \frac{\sqrt{\dma}}{2} \, 
\left( 
\bad 
0 & 0 & 0 \\[0.3cm]
\cdot & a & b \\[0.3cm]
\cdot & \cdot & d 
\ea 
\right) ~, 
\ee
where $a,b,d$ are in general complex parameters of order one. 
Introducing small breaking terms in the first row of $m_\nu$ 
leads to a normal hierarchy, 
i.e., small neutrino masses $m_3 \gg m_2 > m_1$ and a 
very small effective mass of order $\sqrt{\dms} \simeq 10^{-3}$ eV, 
where $\dms$ denotes the mass squared difference of solar neutrinos. 
The common mass scale of the matrix (\ref{eq:le}) is the square root the  
mass squared difference of atmospheric neutrinos $\dma$; 

\item conservation of $L_e - L_\mu - L_\tau$ demands the following form of 
the mass matrix \cite{lelmlt0}: 
\be \label{eq:lelmlt}
m_\nu =  \sqrt{\dma} \, 
\left( 
\bad 
0 & a & b \\[0.3cm]
\cdot & 0 & 0 \\[0.3cm]
\cdot & \cdot & 0 
\ea 
\right) ~.
\ee
The result is an inverted hierarchy, 
i.e., $m_2 \gs m_1 \gg m_3$ and an effective mass of order 
$\sqrt{\dma} \simeq 0.04$ eV.  
Contributions from the charged lepton sector 
are crucial in order to reach accordance with the data \cite{lelmlt1}, 
because solar neutrino mixing is predicted to be maximal 
by Eq.\ (\ref{eq:lelmlt}). Since the deviation from maximal neutrino 
mixing is of order of the Cabibbo angle \cite{ich}, matrices conserving 
$L_e - L_\mu - L_\tau$ are ideal candidates for incorporating 
Quark--Lepton--Complementarity (QLC) \cite{QLC};   

\item conservation of $L_\mu - L_\tau$ has caught only little attention. 
The mass matrix implied by this symmetry has the form \cite{lmlt_o,A4,CR}: 
\be \label{eq:lmlt}
m_\nu =  m_0 \, 
\left( 
\bad 
a & 0 & 0 \\[0.3cm]
\cdot & 0 & b \\[0.3cm]
\cdot & \cdot & 0 
\ea 
\right) ~.
\ee
One is lead to quasi--degenerate light neutrinos, i.e., masses 
much larger than the 
mass splitting, and the effective mass is of order of the common mass scale 
$m_0 \ls $ eV. 

\end{itemize} 
The other possibilities, i.e., conservation of 
$L_\mu$, $L_\tau$, or $L_e - L_\mu$, $L_e - L_\tau$ or 
$L_e + L_\mu - L_\tau$, $L_e - L_\mu + L_\tau$ are ruled out 
phenomenologically. 
Finally, there is a tenth possibility corresponding to 
the conservation of the total lepton charge $L_e + L_\mu + L_\tau$. 
The consequence would be that neutrinos are Dirac particles and that 
there is no neutrinoless double beta decay. 
Without more theoretical input, other than just $L_e + L_\mu + L_\tau$, 
the neutrino mass spectrum and the mixing angles are 
arbitrary (for some recent 
models for Dirac neutrinos, see for instance \cite{dirac}).\\ 

\noindent 
As already mentioned, 
the probably most puzzling aspect of neutrino mixing is that  
one angle is close to zero and another one close to maximal. 
This indicates the presence of a $\mu$--$\tau$ exchange 
symmetry \cite{mutau}, which results in a mass matrix of the form 
\be \label{eq:mutau}
m_\nu = 
\left( 
\bad 
A & B & B \\[0.3cm]
\cdot & D & E \\[0.3cm]
\cdot & \cdot & D 
\ea 
\right)~. 
\ee
By comparing this matrix with the three matrices from 
Eqs.\ (\ref{eq:le}, \ref{eq:lelmlt}, \ref{eq:lmlt}) we see that only 
$L_\mu - L_\tau$ automatically possesses a $\mu$--$\tau$ symmetry. 
Therefore, $\theta_{23} = \pi/4$ and $U_{e3} = 0$ are generic predictions 
of mass matrices obeying the flavor symmetry $L_\mu - L_\tau$.  
Phenomenologically, $\tan \theta_{23}$ is a priori not equal to one 
for (broken) $L_e$ and $L_e - L_\mu - L_\tau$, but just a ratio of 
two numbers. 

\noindent
On the more formal side, it turns out that the combination 
$L_\mu - L_\tau$ is anomaly free and can be 
gauged in the Standard Model (SM) \cite{gauge,gauge1}. 
In fact, also the combination $L_e - L_\mu$ or $L_e - L_\tau$ could be 
chosen to be gauged. However, (as indicated above) these two combinations 
are phenomenologically ruled out, since they typically predict  
two small neutrino mixing angles \cite{gauge1}.\\

\noindent 
Here we wish to analyze the consequences of a see--saw model 
based on the $U(1)$ flavor symmetry $L_\mu - L_\tau$. 
We will break the symmetry in a minimal way so that low energy 
neutrino data can be accommodated and that leptonic $CP$ is possible. 
Interesting and testable phenomenology is predicted. 
Since neutrino physics has entered a precision era, a full study of any 
model has to take radiative corrections 
\cite{RGMatrix,RG2,RG_masses,RG} into account. 
In the framework of $U(1)$ models under consideration, one 
should note the following: models based on $L_e$ receive only minor 
corrections due to RG effects because these are in general very 
small for the normal hierarchy. If $L_e - L_\mu - L_\tau$ is the 
symmetry from which a model originates, then the opposite $CP$ parities 
of the leading mass states suppress the running, thus 
leading again to small effects.  
However, for quasi--degenerate neutrinos, thus in particular 
for $L_\mu - L_\tau$, one expects almost certainly large RG corrections. 
Moreover, a full study of RG effects in a see--saw model 
should not only consider the running from the lowest heavy neutrino mass 
$M_1$ down to the weak scale, but also the running between the different 
heavy neutrino scales $M_3$ to $M_2$ and $M_2$ to $M_1$ 
\cite{RG_oth,RG_see--saw,RGbetween}. Those contributions can be 
at least as important as the usual ones from $M_1$ to $M_Z$. 
In our model we explicitely show that they are. 



\section{\label{sec:model}See--saw Model for  $L_\mu - L_\tau$ } 
\begin{table}[t]
\begin{center}
\begin{tabular}{|c|c|c|c|c|c|c|c|} 
\hline
& $(\nu_e, e)_L$ &  $(\nu_\mu, \mu)_L$ &  
$(\nu_\tau, \tau)_L$ & $N_1, e_R$ & $N_2, \mu_R$ & $N_3, \tau_R$ & $\Phi$ 
\\ \hline \hline 
$L_\mu - L_\tau$ & 0 & 1 & $-1$ & 0 & 1 & $-1$ & 0 \\ \hline
\end{tabular}
\caption{\label{tab:charges}Particle content and charge under the $U(1)$ 
symmetry corresponding to $L_\mu - L_\tau$.}
\end{center}
\end{table}
 
Recently the case $L_\mu - L_\tau$ has been re--analyzed and a simple see--saw 
model has been constructed \cite{CR}. 
Apart from the Standard Model (SM) particle content there are only three 
extra SM singlets required. Those are heavy Majorana neutrinos 
which are used to 
generate light neutrino masses via the see--saw mechanism \cite{see--saw}.
The charge assignment of the particles under $L_\mu - L_\tau$ is 
given in Table \ref{tab:charges}. 
As a consequence, the charged lepton mass matrix is real and diagonal and 
the Lagrangian responsible for the neutrino masses reads 
\bea \label{eq:L} 
-{\cal{L}} = \overline{\nu}_\alpha \, (m_D)_{\alpha i} \, N_i + 
\frac{1}{2} N_i^T C^{-1} \, (M_R)_{ij} \, N_j + h.c. \\[0.3cm] 
= \Phi \, 
\left( 
a \, \overline{\nu}_e  N_1 + b \, \overline{\nu}_\mu  N_2 + 
d  \, \overline{\nu}_\tau N_3 
\right) + 
\frac{\D M}{\D 2} 
\, \left(  X 
\, N_1^T C^{-1} N_1 + 
Y 
\, N_2^T C^{-1} N_3 \right) + h.c.
\eea
Here, $M$ is the high mass scale of the heavy singlets, 
$\Phi$ the Higgs doublet with its vacuum expectation value 
$v/\sqrt{2} \simeq 174$ GeV and  
and $C$ is the charge conjugation matrix. 
In terms of mass matrices, we have 
\be \label{eq:mdMR}
m_D = v \, 
\left( 
\bad 
a & 0 & 0 \\[0.2cm]
0 & b & 0 \\[0.2cm]
0 & 0 & d 
\ea 
\right)
\mbox{ and } 
M_R = M \, 
\left( 
\bad 
X 
& 0 & 0 \\[0.2cm]
\cdot & 0 & Y 
\\[0.2cm]
\cdot & \cdot & 0 
\ea 
\right)~. 
\ee
At this stage, we are dealing with real parameters $a,b,d,X,Y$. 
After integrating out the heavy singlet states, the neutrino mass matrix 
$m_\nu =  - m_D \, M_R^{-1} \, m_D^T$ is given by 
\be \label{eq:mnu0}
m_\nu = - \frac{v^2}{M} 
\left(
\bad 
\D \frac{a^2
}{X} & \D 0 & \D 0 \\[0.3cm]\D
\D \cdot & \D 0 & \D \frac{b \, d
}{Y}  \\[0.3cm]
\D \cdot & \D \cdot & \D 0 
\ea 
\right) ~.
\ee
We need to break $L_\mu - L_\tau$ in order to generate successful 
phenomenology\footnote{For alternative breaking scenarios in 
case of $CP$ conservation, see \cite{CR}.}. 
Note that the mass matrix from above has maximal 23 mixing, 
arbitrary 12 and zero 13 mixing as well as only one 
non--vanishing $\Delta m^2$, 
given by $v^4/M^2~(b^2 d^2/Y^2 - a^4/X^2 ) $. This is the mass 
splitting associated with the solar neutrino oscillations.  
We consider from now on the situation 
in which the symmetry is softly broken by additional small 
parameters in $M_R$. Moreover, in 
contrast to Ref.\ \cite{CR}, we allow for $CP$ violation. 
Since the Dirac and charged lepton 
mass matrices are already diagonal, we can eliminate 
the possible phases of $a,b,d$ by redefining the lepton doublets. 
Consequently, prior to breaking only two phases are present, which we 
assign to the heavy singlets such that we replace in Eq.\ (\ref{eq:mdMR}) 
$X$ with $X \, e^{i \phi}$ and $Y$ with $Y \, e^{i \omega}$. 
The first and minimal approach is to add just one small entry to 
$M_R$.  For instance, we can add to the 12 element an entry 
$\epsilon \, e^{i \chi}$ with real $\epsilon \ll 1$.  
As a result, the low energy mass matrix will read 
\be \label{eq:mnu11}
m_\nu = \frac{\D v^2}{\D M}
\left( 
\bad 
\frac{\D a^2 \, e^{-i \phi}}{\D X}  & 0 & 
\frac{\D -a\, d\, \epsilon\, e^{i(\chi - \omega - \phi)}}{\D X \, Y} \\[0.3cm]
\cdot & 0 & \frac{\D b \, d \, e^{-i \omega}}{\D Y} \\[0.3cm] 
\cdot & \cdot & 
\frac{\D d^2 \, \epsilon^2 \, e^{i(2\chi - 2\omega - \phi)}}{\D X \, Y^2} 
\ea 
\right)~.  
\ee
It is interesting to note that there is no $CP$ violation in 
oscillation experiments or at high energy. This can be shown, for instance, 
by rephasing the elements of Eq.\ (\ref{eq:mnu11}) or by 
considering basis invariant quantities. Let us define here 
the quantity $J_{CP}$, to which any $CP$ violating effect in 
neutrino oscillation experiments will be proportional. 
In terms of mixing angles and the Dirac phase $\delta$ it is given by 
\be \label{eq:jcp0}
J_{CP} = {\rm Im} \{ U_{e1} \, U_{\mu 2} \, U_{e2}^\ast \, U_{\mu 1}^\ast \} 
= \frac{1}{8} \, \sin 2 \theta_{12}\, \sin 2 \theta_{23}\, 
\sin 2 \theta_{13}\, \cos\theta_{13}\, \sin\delta~. 
\ee
From the mass matrix $m_\nu$, in turn, one can deduce \cite{branco} 
\be \label{eq:jcp}
{\rm Im} \left\{ h_{12} \, h_{23} \, h_{31} \right\} = 
\Delta m^2_{21} \, \Delta m^2_{31} \, \Delta m^2_{32}~J_{CP}~, \mbox{ where } 
h = m_\nu^\dagger\, m_\nu~. 
\ee
Using the mass matrix from Eq.\ (\ref{eq:mnu11}) to evaluate this expression 
yields that $J_{CP}$ is zero.

\noindent 
In order to have $CP$ violation, we are therefore forced 
to add another perturbation to $M_R$: 
\be 
M_R = M 
\begin{pmatrix}
X \, e^{i \phi} & \epsilon_1 \, e^{i \psi_1} & 0 \\ 
\cdot  & \epsilon_2 \, e^{i \psi_2}& Y \, e^{i \omega} \\
\cdot  & \cdot & 0 \\
\end{pmatrix}   ~.
\ee
However, one can show that there is only one physical phase, and it 
therefore suffices to consider 
 \begin{equation}
  M_R=M\left(\begin{array}{ccc}
      X  & \epsilon_1 & 0 \\
      \cdot & \epsilon_2 \, e^{i \varphi }& Y \\
      \cdot & \cdot & 0\\
      \end{array}\right)\; 
\end{equation}
to obtain the following low energy effective mass matrix
\be \label{eq:mnufinal}
m_\nu = -\frac{v^2}{M} \, \left(\begin{array}{ccc}
\frac{\D a^2}{\D X} & 0 & \frac{\D -a \, d \, \epsilon_1 }{\D X\, Y}\\[0.3cm]
\cdot & 0 & \frac{\D b \, d }{\D Y}\\[0.3cm]
\cdot &  \cdot & 
\frac{\D -d^2 \left( X \, \epsilon_2 \, e^{i \varphi}-\epsilon_1^2 \right)}
{\D X\,  Y^2}
\end{array}\right)~ .
\ee 
For the rest of this contribution we will focus on this mass matrix. 
Let us therefore analyze in detail the implied phenomenology. 

\noindent 
A matrix with zero entries in the $e\mu$ and $\mu \mu$ elements has 
well--known predictions \cite{2zeros0,2zeros1}. 
In particular, only quasi--degenerate light neutrinos are compatible 
with such a matrix and in addition it is required that the 
$ee$ and the $\mu\tau$ elements are leading and of similar magnitude 
\cite{2zeros1}. This is however just the approximate form of a 
mass matrix conserving $L_\mu - L_\tau$. 

\noindent 
By using the definition of the mass matrix from Eq.\ (\ref{eq:def}) 
and inserting the conditions $m_{e \mu} = 0$ and $m_{\mu \mu} = 0$, 
one can obtain an expression for the ratio of the neutrino masses. 
Expanding in terms of the small parameter $|U_{e3}|$, one finds: 
\bea \label{eq:NuMasses}
  \left|\frac{\D m_1}{\D m_3}\right| 
 \simeq 
\tan^2\theta_{23} - 
|U_{e3}|\, \cos\delta \, \cot\theta_{12} \, 
\frac{\D \tan\theta_{23}}{\D \cos^2\theta_{23}}~, \\[0.3cm]
\left|\frac{\D m_2}{\D m_3}\right| \simeq \tan^2\theta_{23} + |U_{e3}| \, 
\cos\delta \, \tan\theta_{12} \, 
\frac{\D \tan\theta_{23}}{\D \cos^2\theta_{23}}~.
\eea
Since $\theta_{23}$ is close to maximal and $\theta_{13}$ is small,  
the light neutrinos are obviously quasi--degenerate. 
From the ratio of masses one can calculate the ratio of the solar 
mass squared difference over the atmospheric one:  
\begin{equation} \label{eq:R}
  R \equiv \frac{\dms}{\dma} = \frac{m_2^2 - m_1^2}{|m_3^2 - m_2^2|} 
\simeq \left| 4 \, 
  |U_{e3}| \, \cos\delta \, 
\frac{\tan\theta_{23}}{\cos2\theta_{23}} 
\frac{\sin^2\theta_{23}}{\sin2\theta_{12}} \right|\; .
\end{equation}
As $R$ is inverse proportional to the rather small quantity 
$\cos2\theta_{23}$, it is necessary that 
$\sin\theta_{13}\, \cos\delta\, = \mathrm{Re}\,U_{e3} \ll 1$. 
Hence, the Dirac phase should be located around its maximal value 
$\pi/2$, i.e., $CP$ violation is close to maximal. 
The larger $\sin\theta_{13}$ is, the smaller $\cos \delta$ has to be, 
which indicates that for sizable $|U_{e3}|$ only maximal $CP$ violation 
will occur. 
Furthermore, to keep $R$ small, the angle $\theta_{23}$ can not 
become maximal. If it is 
located above $\pi/4$, then it follows from Eq.\ (\ref{eq:NuMasses}) 
that $m_1 > m_3$ and $m_2 > m_3$, which leads to an inverted ordering. 
Equivalently, for $\theta_{23} < \pi/4$ one has 
$m_3 > m_1$ and $m_3 > m_2$ and 
the mass ordering is normal. 
These aspects of the phenomenology are almost independent on the precise value 
of $\theta_{12}$, which --- as we will see later on --- can receive 
large renormalization corrections. 
The effective mass governing 
neutrinoless double beta decay can be written as 
\begin{equation}
\meff = \frac{v^2}{M} \, \frac{a^2}{X} \simeq \tan^2\theta_{23} 
\sqrt{\frac{\dma }{\left|1-\tan^4\theta_{23}\right|}}~,  
\end{equation}
which shows again that maximal atmospheric neutrino mixing 
is forbidden. 
For later use, we note here that for the 
neutrino texture under discussion typically 
$\varphi_1 \simeq \varphi_2 \simeq \pi$  
holds, i.e., the two Majorana phases 
have very similar values \cite{2zeros1}. This will be very useful 
when we analyze the renormalization aspects of the model. 

\begin{figure}[b]\vspace{-3cm}
\begin{center}\vspace{-3cm}
\hspace{-3cm}
\epsfig{file=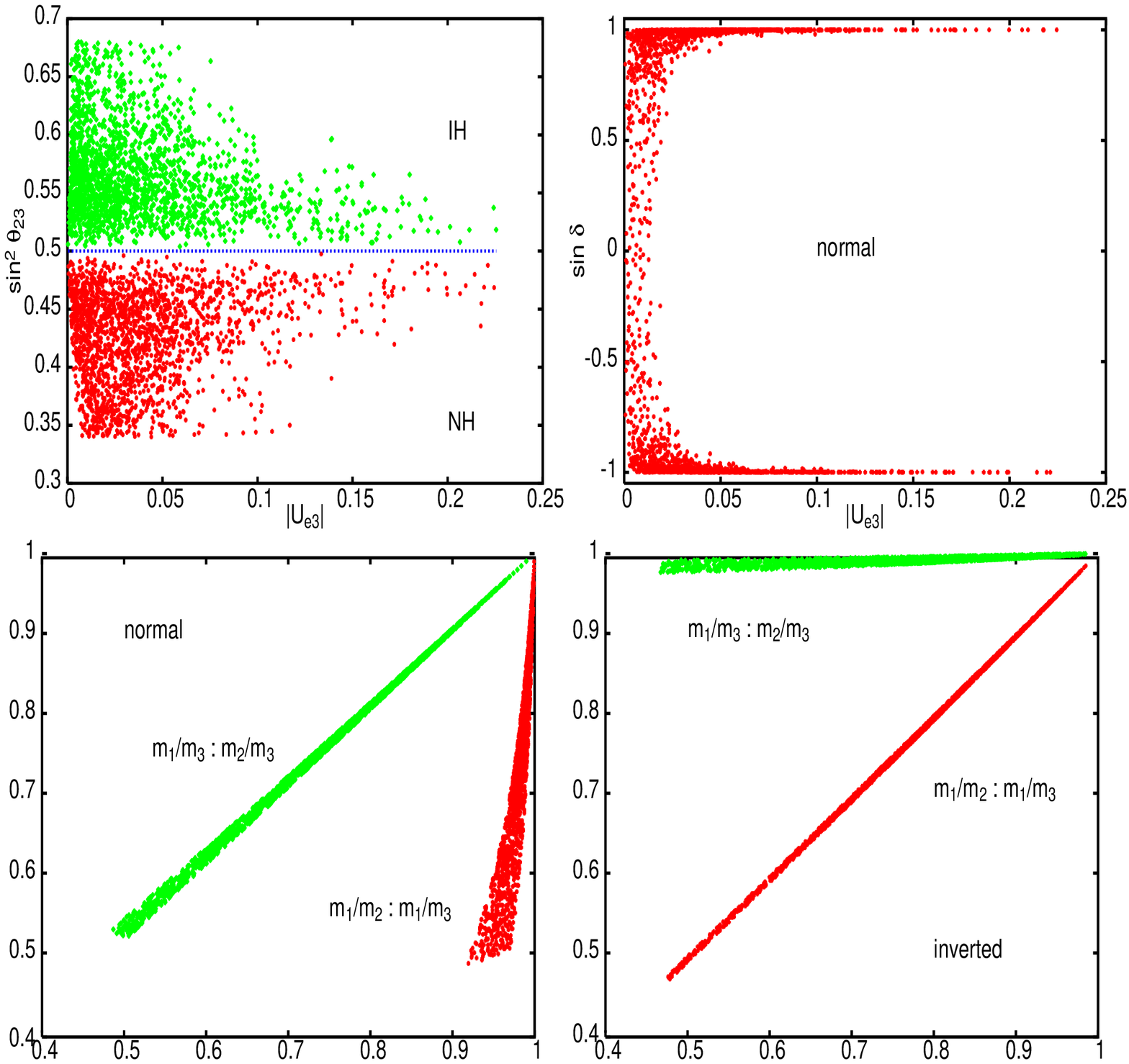,width=19cm,height=24cm}
\vspace{-4cm}
\caption{\label{fig:zero1222}Scatter plots of some 
of the observables resulting from a mass matrix with zero 
$e \mu$ and $\mu\mu$ elements. We required the $3\sigma$ values for the 
oscillation parameters from \cite{valle}.}
\end{center}
\end{figure}

\noindent 
Let us now take a closer look at the structure of the Dirac and 
heavy Majorana mass matrices. 
Since the charged leptons display a hierarchy, it is natural to assume 
that also the eigenvalues of the Dirac mass matrix are hierarchical. 
Then it is required that also the heavy Majorana neutrinos 
display a hierarchy in the form of $Y \gg X$. 
Typical values of the parameters which in this case successfully reproduce 
the neutrino data are 
$Y={\cal O}(1)$, $a={\cal O}(0.01)$, $ b \sim d={\cal O}(0.1)$, 
$X={\cal O}(0.001)$, $\epsilon_1={\cal O}(0.001)$ and  
$\epsilon_2={\cal O}(0.1)$. With these values, 
the eigenvalues of $M_R$ are approximatively given by $M \, X$ and 
$M\left(Y \pm \epsilon_2/2\right)$. 
We plot in Fig.\ \ref{fig:zero1222} some of the correlations obtained for 
parameters around the values given above. The oscillation parameters are 
required to lie within the $3\sigma$ ranges from Ref.\ \cite{valle}.
In fact, just demanding zero 
entries in the $e\mu$ and $\mu\mu$ elements will lead to such 
a behavior of the observables \cite{2zeros1}. 
Indeed, as can be seen from the Figure, the neutrinos are close in mass, 
atmospheric neutrino mixing is not maximal 
and lies above (below) $\pi/4$ for the inverted (normal) 
mass ordering. Furthermore, $CP$ violation is basically maximal 
for sizable values of $|U_{e3}|$.\\

\section{\label{sec:RG}Renormalization Group Effects}
As the flavor symmetry $L_\mu-L_\tau$ leads to 
quasi--degenerate neutrino masses, strong
running of the mixing angles is generically expected \cite{RG2,RG}.  
Typically, the running of the mixing angles $\theta_{ij}$ 
in a quasi--degenerate mass scheme with a common mass scale $m_0$ is 
proportional to $m_0^2/(m_i^2 - m_j^2)$ and therefore 
particularly strong for $\theta_{12}$. 
This behavior turns out to hold not only for the running between 
the scale of the lightest heavy Majorana neutrino and the weak scale, 
but also when the running between the see--saw scales is taken into 
account \cite{RG_oth,RG_see--saw}. 
The heavy Majorana neutrinos have to be integrated out one after the other, 
leading to a series of effective theories. It turns out that the running 
of the observables 
between the scales $M_3$ and $M_2$ or $M_2$ and $M_1$ can be of the 
same order as the running between $M_1$ and $M_Z$. In general 
this leads to quite involved expressions for the $\beta$--functions. 
However, in our case the structure of the Dirac and charged 
lepton mass matrices (i.e., the fact that they are diagonal) 
simplifies matters considerably and allows for some analytic 
understanding of the numerical results. 
All of the expressions we give in this Section are understood to be 
precise to order $|U_{e3}|$ and can be obtained with the help of Refs.\ 
\cite{RG,RG_see--saw}. 
For instance, the $\beta$--functions for the mixing angles are:  
\begin{align} \label{eq:beta_mix}
  16\pi^2 \, \dot\theta_{12} \nonumber 
  \simeq&\frac{m_0^2}{\dms}
\left(1+\cos\left(\varphi_2-\varphi_1\right)\right)\, \sin2\theta_{12}\, 
\left[P_{11}-\left(P_{22}\, \cos^2\theta_{23}+P_{33}\, \sin^2\theta_{23}\right)
\right]~,\\
   32\pi^2 \, \dot\theta_{13}
   \simeq&\frac{m_0^2}{\dma}
\left(\cos\left(\delta-\varphi_1\right)-\cos\left(\delta-\varphi_2\right)
\right) \, \sin 2\theta_{12} \, \sin2\theta_{23} \, 
\left(P_{22}-P_{33}\right)~,\\ \nonumber 
   16\pi^2 \, \dot\theta_{23}
   \simeq&\frac{m_0^2}{\dma}
\left[\left(1+\cos\varphi_2\right)\, \cos^2\theta_{12}+
\left(1+\cos\varphi_1\right)\, \sin^2\theta_{12}\right]\, 
\sin2\theta_{23}\, \left(P_{22}-P_{33}\right)~.
\end{align}
Note that since 
$\varphi_1 \simeq \varphi_2$ there is no cancellation in the 
first relation for $\dot\theta_{12}$. 
The matrix $P$ in the above equations is given in the SM and in the 
Minimal Supersymmetric Standard Model (MSSM) by 
\begin{equation}
  P=\left\{
    \begin{array}{cl}
        -\frac{3}{2} Y_e Y_e^\dagger + 
\frac{1}{2} Y_\nu Y_\nu^\dagger& \mathrm{SM}~, \\[0.2cm]
        Y_e Y_e^\dagger + Y_\nu Y_\nu^\dagger& \mathrm{MSSM}~.\\
      \end{array}\right. 
  \end{equation}
Since in our model both $Y_\nu = m_D/v$ and $Y_e = m_\ell/v$ are 
diagonal, $P$ is diagonal as well and no off--diagonal elements appear 
in Eq.\ (\ref{eq:beta_mix}). 
With the 
values of the parameters in $m_D$ as given in Section 
\ref{sec:model} (i.e., $a \ll b \sim d = {\cal O}(0.1)$) 
we can safely neglect $P_{11}$.  
This leads in particular to a negative $\beta$--function 
for $\theta_{12}$. 
Hence, this angle will become larger when evolved from high to 
low scales. 
Neglecting further the charged lepton Yukawas in $Y_e$ above the see--saw scales
and noting 
that $P_{22} \simeq b^2 /2$ and $P_{33} \simeq d^2 /2$ for the SM 
and twice those values for the MSSM, we see that the running of 
$\theta_{13}$ and $\theta_{23}$ is suppressed with respect to the 
running of $\theta_{12}$ due to two reasons: 
firstly, it is inverse proportional to \dma{} and secondly, it is 
proportional to 
$P_{22} - P_{33} \propto b^2 - d^2$, which is smaller than 
$P_{22} + P_{33}$, on which the running of $\theta_{12}$ (approximately) 
depends. 
Hence, the running of $\theta_{13}$ and $\theta_{23}$ is suppressed by 
$R~(b^2 - d^2)/b^2$ above the see--saw scales. 

\noindent 
Below the see--saw scales only the tau--lepton Yukawa coupling $y_\tau$ 
governs the RG corrections. In this regime the evolution 
goes like \cite{RG}
\be
16 \, \pi^2 \, \dot\theta_{12} \simeq -y_\tau^2 \, \sin 2 \theta_{12} \, 
\sin^2 \theta_{23} \, \frac{m_0^2}{\dms} 
\left( 1 + \cos (\varphi_2 - \varphi_1 )\right) ~, 
\ee
in the MSSM, which is again negative and leads to sizable running. 
The formulae for the running of $\theta_{13,23}$ are again suppressed by 
roughly a factor $R$. 

\noindent 
The phases stay almost constant in the whole range, 
because one can show that for our model 
$\Dot\varphi_1$ and $\Dot\varphi_2$ are mainly proportional 
to $\sin \varphi_1$ and $\sin\varphi_2$, respectively.  
Roughly the same behavior is found below $M_1$. 
Finally, the RG effects on the neutrino masses correspond predominantly to 
a rescaling, since the flavor--diagonal couplings, 
i.e., gauge couplings and the quartic Higgs coupling, dominate the evolution 
\cite{RG_masses}.\\

\noindent 
We can analyze if the zero entries 
in the mass matrix Eq.\ (\ref{eq:mnufinal}) remain zero entries. 
Below the see--saw scale it is well known that the RG corrections are 
multiplicative on the mass matrix, a fact which leaves zero entries zero. 
Taking the running in between the heavy Majorana masses into account, 
one can factorize the renormalization group  
effects $Z_\mathrm{ext}$ from the tree--level neutrino mass matrix 
in the MSSM as 
\cite{RGbetween}
\begin{equation}
  m_\nu= Z_\mathrm{ext}^T \, m_\nu^0 \, Z_\mathrm{ext}\; . 
\end{equation}
Since in our model the 
neutrino and the charged lepton Yukawa coupling matrices are diagonal, 
the renormalization group effects are flavour diagonal, too. 
Therefore, texture zeroes in the charged lepton basis remain zero. 
With the already mentioned simplifications, 
$Z_\mathrm{ext}$ is approximately\footnote{As the perturbations are small, 
the mass eigenvalues of the heavy right--handed neutrinos are 
approximately given by 
$\left(M_1,M_2,M_3\right) \simeq 
M\left(X,Y(1-\epsilon_2/2),Y(1+\epsilon_2/2)\right) \simeq 
\left(X,Y,Y\right)$.} given by 
 \begin{multline}
   16\pi^2 \left(Z^\mathrm{MSSM}_\mathrm{ext}-\mathbbm{1}\right)=
\diag\left(0,\, b^2\ln\frac{M \, Y}{\Lambda},\, d^2\ln\frac{M \, Y}{\Lambda}+
y_\tau^2\ln\frac{v}{\Lambda}\right)\\
 +b^2\ln\frac{M \, Y }{\Lambda}+d^2\ln\frac{M \, Y}{\Lambda}
 +\left[-\frac{3}{5}
   g_1^2 - 3 g_2^2 + 3 y_t^2\right]\ln\frac{v}{\Lambda}\; ,
 \end{multline}
where $\left(g_2,\,g_1\right)$ are the gauge couplings\footnote{We 
use GUT charge normalization for $g_1$, i.e., 
$g_1^\mathrm{SM}=\sqrt{\frac{3}{5}}g_1$.} of the electroweak 
interactions $SU(2)_L \times U(1)_Y$, 
$y_t$ and $y_\tau$ are the top quark and $\tau$ Yukawa couplings, 
respectively and $\Lambda$ is the GUT scale of $2 \cdot 10^{16}$ GeV.

\noindent 
In the extended SM, however, there are additional 
corrections which can not be factorized. 
They are responsible for the instability of texture zeroes under the
renormalization group in the SM~\cite{RG_see--saw, RGbetween}. 
We checked that for most observables the running behavior 
in the SM is similar to the running in case of the MSSM. 
The solar neutrino mixing angle receives however more RG corrections 
in the SM, a fact which can be traced back to the 
appearance of an $e\mu$ entry in the mass matrix Eq.\ (\ref{eq:mnufinal}). 
One might wonder at this point if this filling of zero entries would 
allow one to generate successful phenomenology from a mass matrix obeying the 
flavor symmetry $L_\mu - L_\tau$ without any breaking, i.e., just 
from Eq.\ (\ref{eq:lmlt}).  
Recall, however, that in the SM $L_\mu - L_\tau$ is anomaly free and 
therefore the texture of the mass matrix is stable. 

\begin{figure}[t]
\begin{center}
\epsfig{file=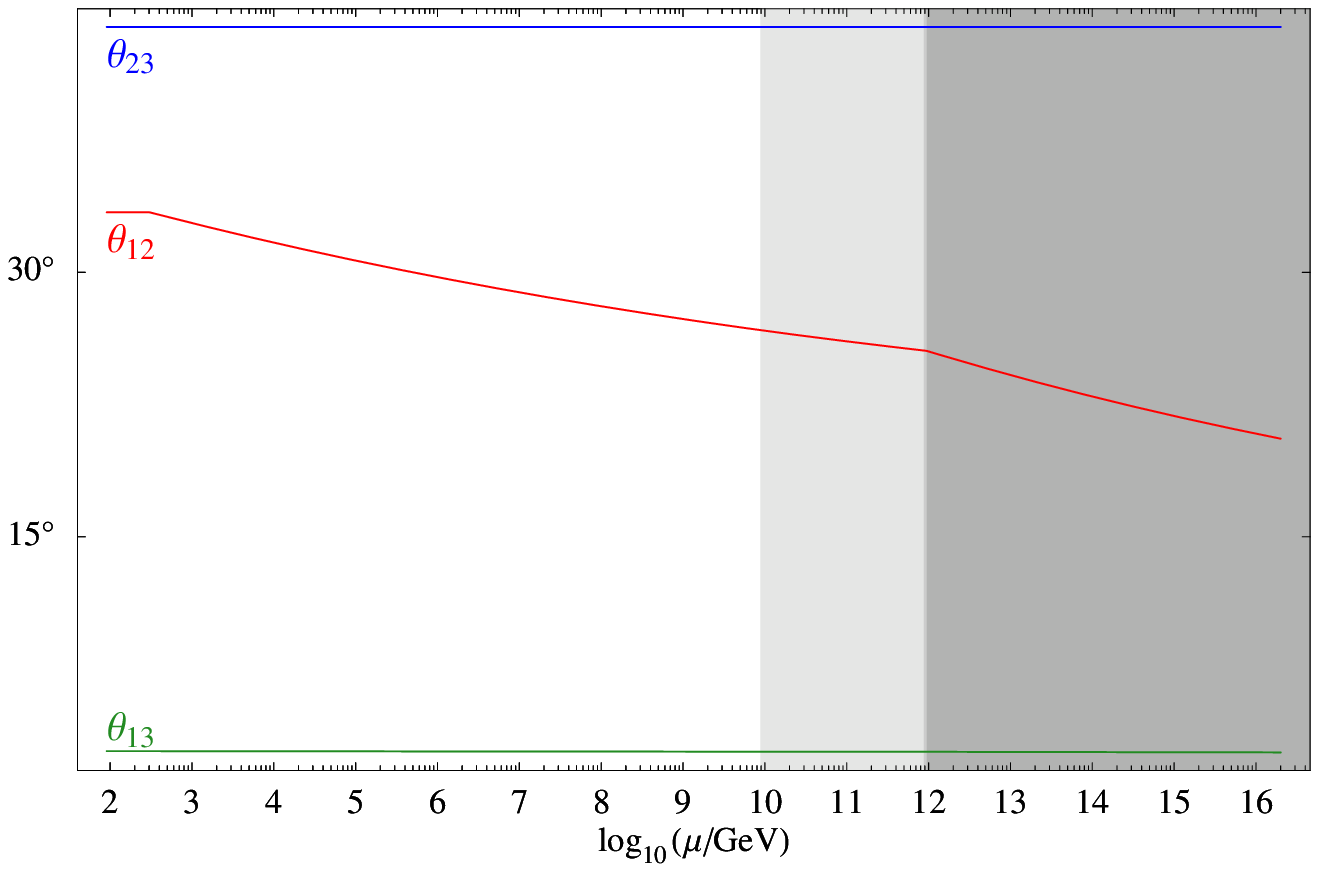,width=7.7cm}
\hspace{.5cm}
\epsfig{file=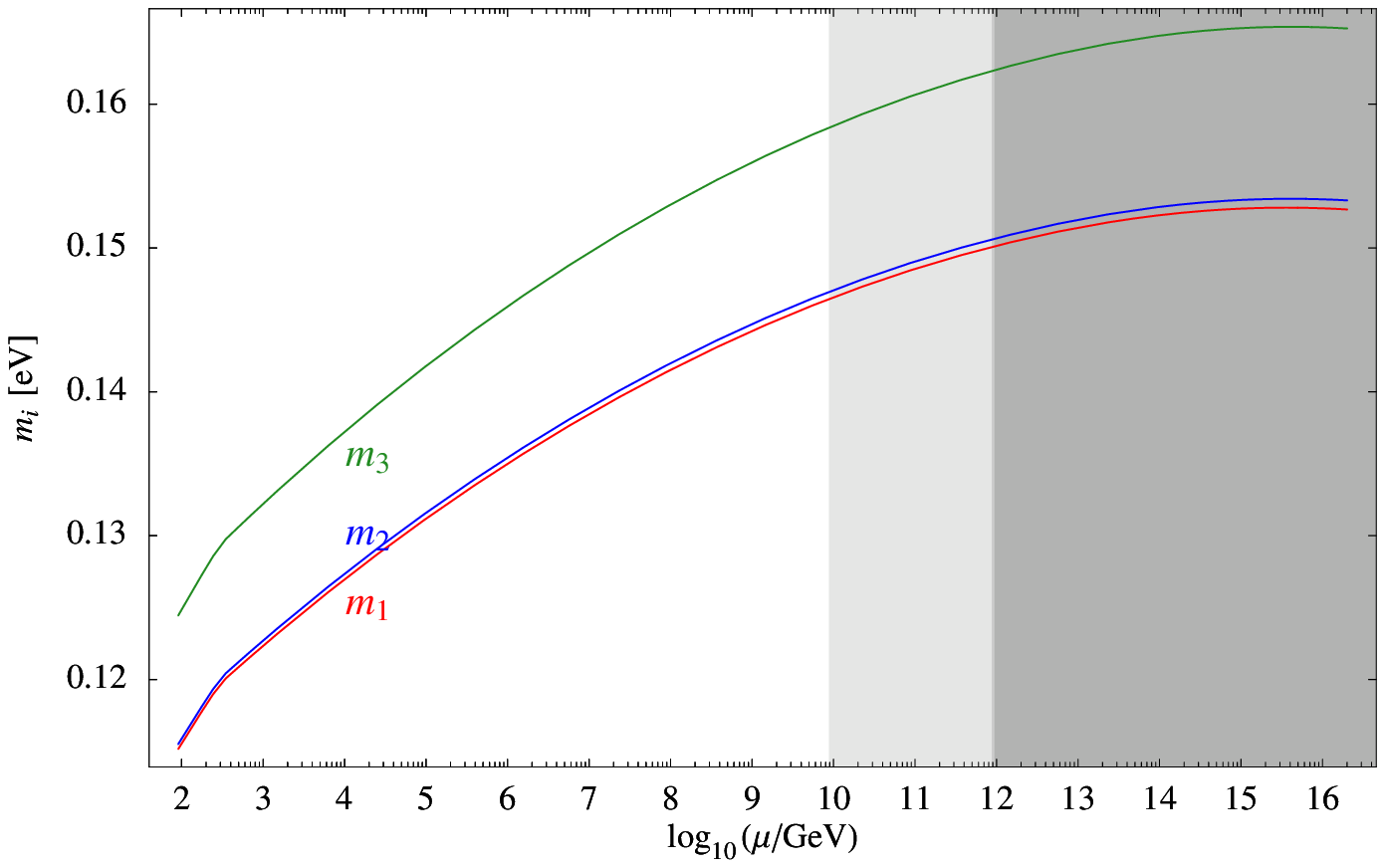,width=7.7cm}

\vspace{.5cm}

\epsfig{file=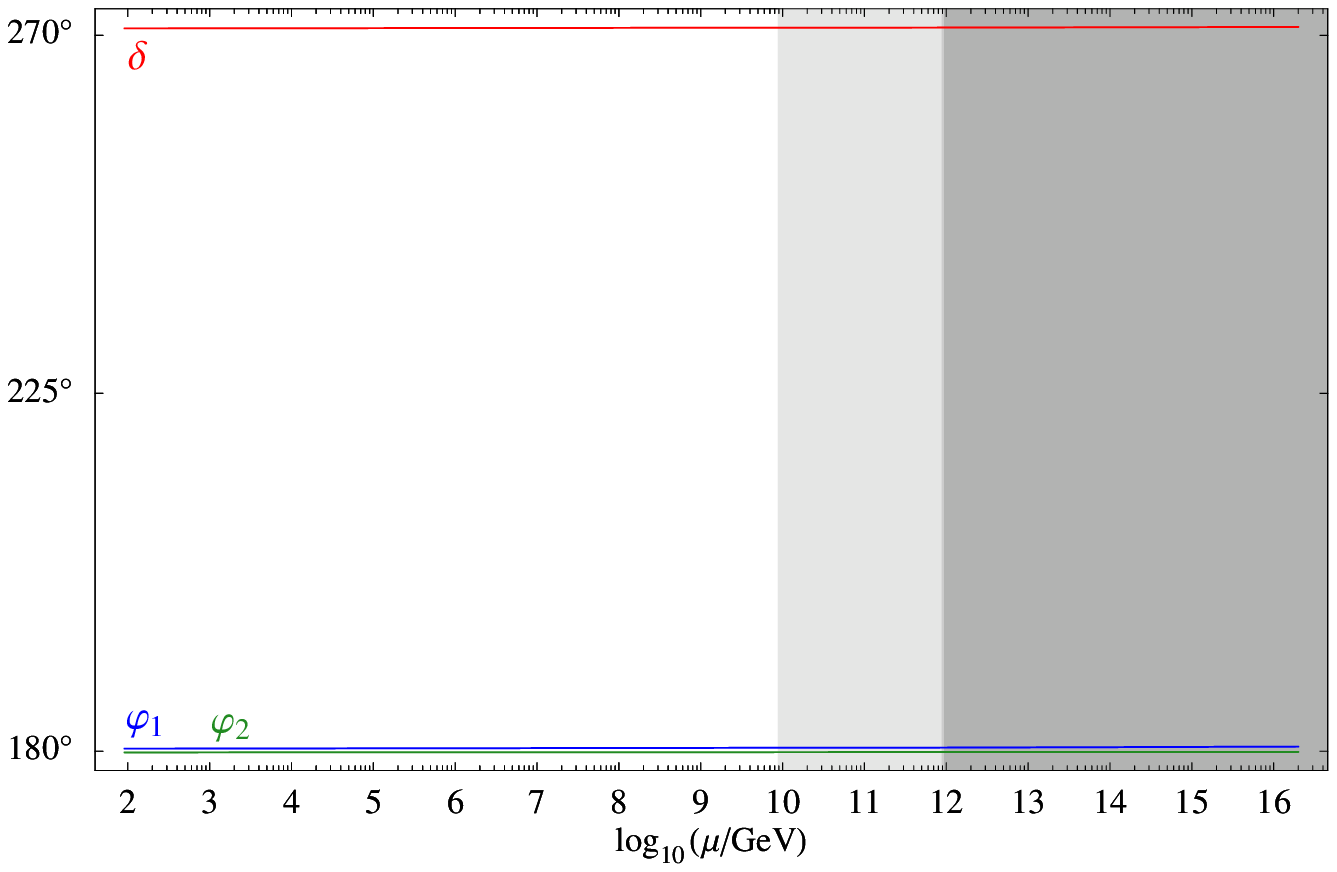,width=7.7cm}
\hspace{.5cm}
\epsfig{file=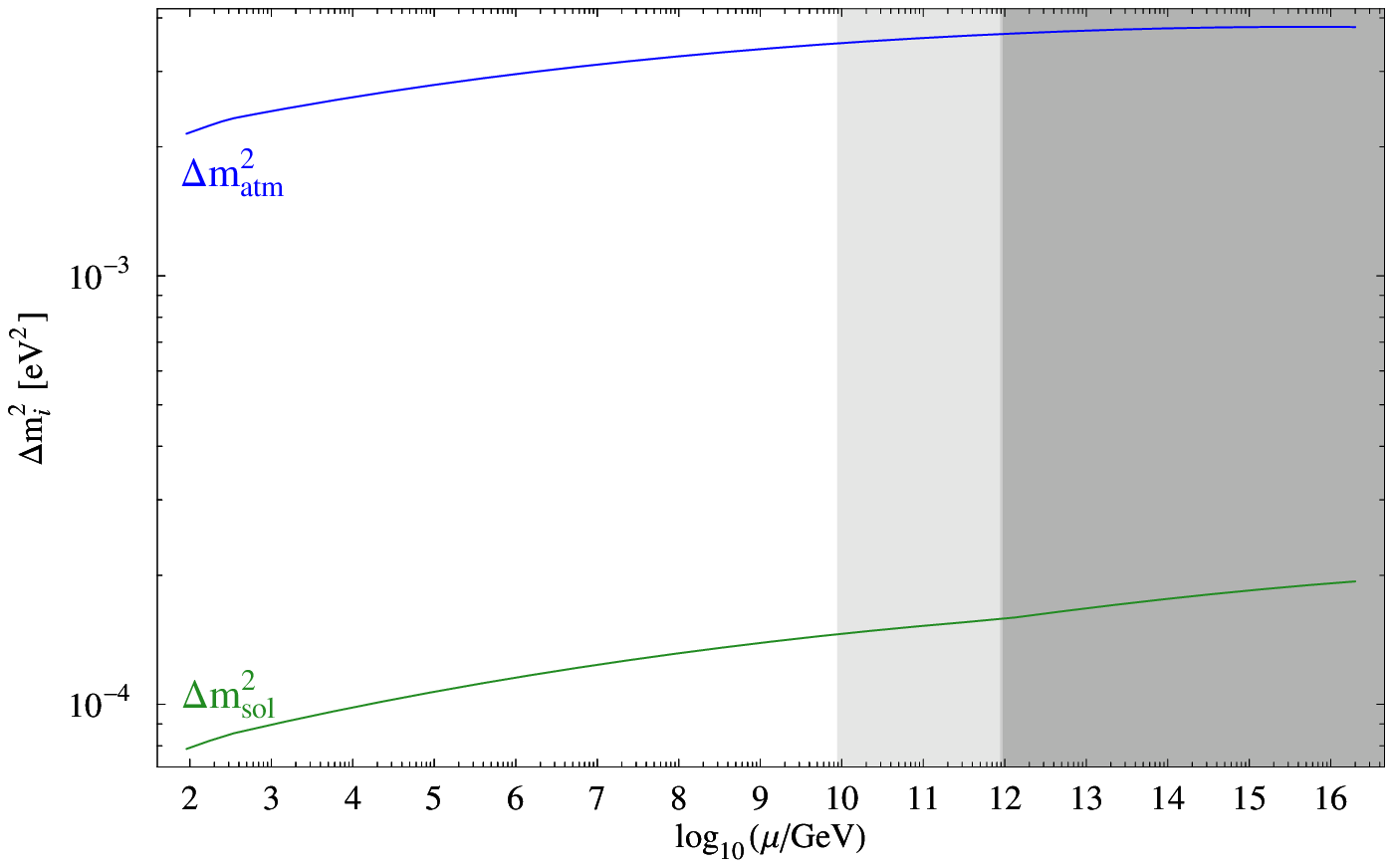,width=7.7cm}
\caption{\label{fig:MSSM}Typical plot for the evolution of the 
mixing angles and masses from the GUT scale $2 \times 10^{16}$ 
GeV to low scale. In this example, we have $\tan \beta=10$ and the 
parameters in the matrix Eq.\ (\ref{eq:mnufinal}) at the GUT scale are
$a=0.0066926$, $b=0.0692883$, $c=0.0697464$, $X=0.0096528$, $Y=1$,
$\epsilon_1=0.0005595$, $\epsilon_2=0.0749098$, $\varphi=2.45376$ and $M =
9.098937108 \cdot 10^{11}$ GeV. }
\end{center}
\end{figure}

\noindent 
We plot in Fig.\ \ref{fig:MSSM} the running of the angles, phases and  
masses for a typical example in the MSSM for $\tan \beta = 10$ \cite{REAP}.  
The resulting neutrino parameters at the GUT
scale are $\sin^2 \theta_{12} = 0.123 $, 
$U_{e3} = 0.0484\, e^{\im 4.73}$, $\sin^2 \theta_{23} = 0.481 $, 
$(m_1, \, m_2, \, m_3) = (0.1527, \, 0.1533, \, 0.1653) \eV$ with $\dms =
1.9 \cdot 10^{-4} \eV^2 $ and 
$\dma = 3.8 \cdot 10^{-3} \eV^2$. They are changed by the renormalization
group evolution to $\sin^2 \theta_{12} = 0.303$, 
$U_{e3} = 0.0496\, e^{\im 4.73}$, 
$\sin^2 \theta_{23} = 0.481$, 
$(m_1, \, m_2, \, m_3) = (0.1152, \, 0.1155, \, 0.1245)\eV$ with $\dms =
7.9 \cdot 10^{-5}\eV^2 $ and $\dma = 2.1 \cdot 10^{-3}\eV^2$. 
Note that the phases and $\theta_{13,23}$ remain practically constant, 
whereas $\sin^2 \theta_{12}$ and $\Delta m^2_{\odot, \rm A}$ are 
changed by factors of up to three, and that the running in between 
and above the 
see--saw scales is at least as important as the running below them. 
This implies that radiative corrections, in particular for 
$\theta_{12}$ and the mass squared differences, can be crucial if one 
wishes to make meaningful phenomenological predictions of a given model.

\begin{table}[t]
\begin{center}
\begin{tabular}{|c|c|c|} 
\hline
$L'$ & matrix & comments \\ \hline \hline
$\ba L_e \\ \mbox{\small normal hierarchy} \\
\mbox{\small always small RG effects} \ea$ & 
$\left( \bad 0 & 0 & 0 \\ \cdot & a & b \\ 
\cdot & \cdot & d \ea \right) $
& $\ba 
R = \frac{\dms}{\dma} \simeq  |U_{e3}|^2 \\
\tan^2 \theta_{\rm atm} \simeq 1 +  |U_{e3}| \simeq 1 + \sqrt{R} \\
{\meff \simeq \sqrt{\dma} \, |U_{e3}|^2 \simeq \sqrt{\dms} } 
\ea$  
\\ \hline 
$\ba L_e - L_\mu - L_\tau \\ \mbox{\small inverted hierarchy} 
\\ \mbox{\small small RG effects (pseudo--Dirac)} 
\ea $ & 
$\left( \bad 0 & a & b \\ \cdot & 0 & 0 \\ \cdot & \cdot & 0 \ea \right) $
& $ \ba \mbox{requires } U_{\ell} \mbox{: ideal for QLC} \\
\ts \simeq 1 - 4 \, |U_{e3}| \stackrel{?}{\simeq} 
1 - 2\sqrt{2} \, {\sin \theta_{\rm C} }\\
{\meff \simeq \sqrt{\dma} }
\ea$ \\ \hline 
$\ba {L_\mu - L_\tau} \\ \mbox{\small quasi--degeneracy} 
\\ \mbox{\small always large RG effects} 
\ea $ & 
$\left( \bad a & 0 & 0 \\ \cdot & 0 & b \\ \cdot & \cdot & 0 \ea \right) $ 
& $ \ba \mbox{in leading order:}\\
U_{e3}=0 \mbox{ and } \theta_{23} = \pi/4 \\
{\meff \simeq m_0 }
\ea $ \\ \hline
\end{tabular}
\caption{\label{tab:sum}The three simple $U(1)$ charges as 
implied by current data, the zeroth order mass matrix and some comments on 
their phenomenology.}
\end{center}
\end{table}

\section{\label{sec:concl}Conclusions and Summary}
Identifying the structure of the mass matrix  $m_\nu$ is one of the main 
goals of neutrino physics. Current data seems to point towards three 
interesting forms of $m_\nu$, each corresponding to the 
(approximative) conservation of a non--standard lepton charge. 
While $L_e$ and $L_e - L_\mu - L_\tau$ are well known cases --- leading to 
the normal and inverted hierarchy, respectively --- 
the third possibility $L_\mu - L_\tau$ has not been examined in as 
much detail. Since quasi--degenerate neutrinos are predicted by the 
conservation of $L_\mu - L_\tau$, this case is easily 
testable. Moreover, since the mass matrix is automatically 
$\mu$--$\tau$ symmetric, 
further phenomenological support is provided. 
We stress again that the three possibilities 
$L_e$, $L_e - L_\mu - L_\tau$ and $L_\mu - L_\tau$ differ in particular 
by their prediction for neutrinoless double beta decay, a fact which will 
easily allow to identify the correct one. In Table \ref{tab:sum} 
the relevant matrices and some aspects of their phenomenology are 
summarized. 
Here we analyzed the consequences of a see--saw model based on 
$L_\mu - L_\tau$, which predicts, e.g., non--maximal atmospheric neutrino 
mixing, large $CP$ violation and quasi--degenerate neutrinos. 
This last aspect makes radiative corrections be particularly 
important and we carefully analyzed their effects both 
numerically and analytically. We showed explicitely that running 
between the see--saw scales is at least as important as running 
from the lowest see--saw scale down to low energies.

\vspace{0.5cm}
\begin{center}
{\bf Acknowledgments}
\end{center}
We thank R.~Foot for discussions and W.R.~wishes to thank S.~Choubey 
for a fruitful collaboration. 
This work was supported by the ``Deutsche Forschungsgemeinschaft'' in the 
``Sonderforschungsbereich 375 f\"ur Astroteilchenphysik'' 
and under project number RO--2516/3--1.

\end{document}